\documentclass[aps,prb,reprint,groupedaddress,preprintnumber,showpacs]{revtex4-1}
\usepackage{graphicx}
\usepackage{color}
\usepackage{dcolumn}

\begin{document}


\title{Multiband effects and the possible Dirac states in LaAgSb$_2$}
\author{Kefeng Wang}
\affiliation{Condensed Matter Physics and Materials Science Department, Brookhaven National Laboratory, Upton New York 11973 USA}
\author{C. Petrovic}
\affiliation{Condensed Matter Physics and Materials Science Department, Brookhaven National Laboratory, Upton New York 11973 USA}

\date{\today}

\begin{abstract}
Here we report the possible signature of Dirac fermions in the magnetoresistance, Hall resistivity and magnetothermopower of LaAgSb$_2$. The opposite sign between Hall resistivity and Seebeck coefficient indicates the multiband effect. Electronic structure calculation reveals the existence of the linear bands and the parabolic bands crossing the Fermi level. The large linear magnetoresistance was attributed to the quantum limit of the possible Dirac fermions or the breakdown of weak-field magnetotransport at the charge density wave phase transition. Analysis of Hall resistivity using two-band model reveals that Dirac holes which dominate the electronic transport have much higher mobility and larger density than conventional electrons. Magnetic field suppresses the apparent Hall carrier density, and also induces the sign change of the Seebeck coefficient from negative to positive. These effects are possibly attributed to the magnetic field suppression of the density of states at the Fermi level originating from the quantum limit of the possible Dirac holes.
\end{abstract}

\maketitle

\section{Introduction}
Recently layered rare-earth antimonides have been attracting wide attention due to their possible relationship to superconductivity, charge density wave (CDW) and colossal magnetoresistance (MR).\cite{lasb1,lasb2,lasb3,lasb4} For example, two-dimensional superconductivity ($T_c\sim1.7$ K) was observed in LaSb$_2$ under ambient pressure, while application of hydrostatic pressure induces a two- to three-dimensional superconducting crossover.\cite{lasb2,lasb3} Large positive magnetoresistance was observed in LaSb$_2$ and LaAgSb$_2$.\cite{lasb3,lasb4,lasb5,laagsb1,laagsb2} LaSb$_2$ exhibits a large linear magnetoresistance with no sign of saturation up to 45 T fields and $(\rho(H)-\rho(0))/\rho(0)\sim 10^4\%$.\cite{lasb4} MR in LaAgSb$_2$ is also linear and reaches $\sim 2500\%$ in 18 T field.\cite{laagsb1} The resistivity of LaAgSb$_2$ exhibits a significant anomaly at $\sim 210$K which was attributed to the possible CDW transition.\cite{laagsb1,laagsb2} Very large magnetothermopower effects were observed in LaAgSb$_2$.\cite{thermopower} The magnetoresistance and magnetothermopower effects in these materials are extraordinary and are still poorly understood. This is because the semiclassical transport in conventional metals gives quadratic field-dependent MR in the low field range which would saturate in the high field and the diffusive Seebeck coefficient does not depend on the external magnetic field.\cite{mr3}

Large linear MR was also observed recently in SrMnBi$_2$ which has similar crystal structure to LaAgSb$_2$.\cite{srmnbi21,srmnbi22,srmnbi23} Crystal structure of SrMnBi$_2$ contains alternatively stacked MnBi layers and two-dimensional Bi square nets separated by Sr atoms along the $c$-axis.\cite{srmnbi21,srmnbi22} Highly anisotropic Dirac states were identified in SrMnBi$_2$ where linear energy dispersion originates from the crossing of two Bi $6p_{x,y}$ bands in the double-sized Bi square net.\cite{srmnbi21} The linear MR is attributed to the quantum limit of the Dirac fermions.\cite{srmnbi21,srmnbi23,quantummr} 
In high enough field and the electronic systems enter the extreme quantum limit where all of the carriers occupy only the lowest Landau level (LL) and a large linear MR could be expected.\cite{quantummr,quantumtransport} Unlike the conventional electron gas with parabolic energy dispersion, the distance between the lowest and $1^{st}$ LLs of Dirac fermions in magnetic field is very large and the quantum limit is easily realized in low field region.\cite{LL1,LL2} Consequently some quantum transport phenomena such as quantum Hall effect and large linear MR could be observed in the low field region for Dirac fermion materials, such as graphene,\cite{LL1,LL2} topological insulators,\cite{qt1,qt2,qt3} Ag$_{2-\delta}$Te/Se,\cite{agte1,agte2,bi} iron-based superconductors BaFe$_2$As$_2$~\cite{qt4,qt5} and La(Fe,Ru)AsO,\cite{lafeaso} as well as SrMnBi$_2$.\cite{srmnbi21,srmnbi23}

In this paper, we attribute the large linear magnetoresistance and magnetothermopower in LaAgSb$_2$ to the quantum limit of Dirac fermions as revealed by ab initio calculation, or the breakdown of weak-field magnetotransport at CDW phase transition. Interestingly, the Hall resistivity is positive, but the Seebeck coefficient is negative in 0 T field. Analysis of Hall resistivity using two-band model reveals that Dirac holes have higher mobility and larger density then conventional electrons, and dominate the electronic transport. Magnetic field suppresses the apparent Hall carrier density, and also induces the sign change of the Seebeck coefficient from negative to positive. These effects are attributed to the magnetic field suppression of the density of states at the Fermi level originating from the quantum limit of the Dirac holes.

\section{Experimental}

Single crystals of LaAgSb$_2$ were grown using a high-temperature self-flux method.\cite{laagsb1} The resultant crystals are plate-like. X-ray diffraction (XRD) data were taken with Cu K$_{\alpha}$ ($\lambda=0.15418$ nm) radiation of Rigaku Miniflex powder diffractometer. Electrical transport measurements up to 9 T were conducted in Quantum Design PPMS-9 with conventional four-wire method. In the in-plane measurements, the current path was in the \textit{ab}-plane, whereas magnetic field was parallel to the \textit{c}-axis. Thermal transport properties were measured in Quantum Design PPMS-9 from 2 K to 350 K using one-heater-two-thermometer method. The direction of heat and electric current transport was along the $ab$-plane of single grain crystals with magnetic field along the \textit{c}-axis and perpendicular to the heat/electrical current. The relative error in our measurement was $\frac{\Delta \kappa}{\kappa}\sim$5$\%$ and $\frac{\Delta S}{S}\sim$5$\%$ based on Ni standard measured under identical conditions. Fist principle electronic structure calculation were performed using experimental lattice parameters within the full-potential linearized augmented plane wave (LAPW) method ~\cite{wien2k1} implemented in WIEN2k package.\cite{wien2k2} The general gradient approximation (GGA) of Perdew \textit{et al}., was used for exchange-correlation potential.\cite{gga} 

\section{Results and discussions}

The crystal structure of LaAgSb$_2$ (Fig. 1(a)) consists of intercalated La ions between alternatively stacked two-dimensional Sb layers (red balls) and AgSb layers along the $c$-axis. The first principle electronic structure calculation reveals that the density of states (DOS) at Fermi level is very small (Fig. 1(b)) and is dominated by the states coming from two-dimensional Sb layers. There are three nearly linear narrow bands crossing Fermi level along $Z-A$, $Z-R$ and $\Gamma-M$ directions respectively (as indicated by the red circles in Fig. 1(c)). Because of the occupation of Ag ions below and above the quasi-two-dimensional Sb layers, the unit cell of Sb layer has two Sb atoms (Fig. 1(a)). This will lead to the folding of the dispersive $p$ orbital of Sb. The two $p_{x,y}$ bands from two Sb atoms cross each other at a single point and then form the nearly linear band and Dirac-cone-like point around the Fermi level (Fig. 1(c)). This is similar to the case of SrMnBi$_2$ where the crossing of Bi 6$p_{x,y}$ orbitals forms the Dirac-cone-like point.\cite{srmnbi21}

Powder X-ray diffraction results confirmed phase purity of our crystals, and the temperature and magnetic field dependent resistivity and Seebeck coefficient agrees very well with previous report.\cite{laagsb1,laagsb2,thermopower}.\cite{laagsb1,laagsb3} Fig. 2 shows the thermal conductivity $\kappa$ in 0 T and 9 T magnetic field perpendicular to $ab$-plane.  The thermal conductivity $\kappa$ exhibits a CDW anomaly at about 210 K (blue dashed line) that corresponds to similar anomalies in $\rho(T)$ and $S(T)$ already reported in literature.\cite{laagsb1,laagsb2,thermopower} It is suppressed significantly by the magnetic field below 200 K (Fig. 2).

\begin{figure}[tbp]
\includegraphics[scale=0.45]{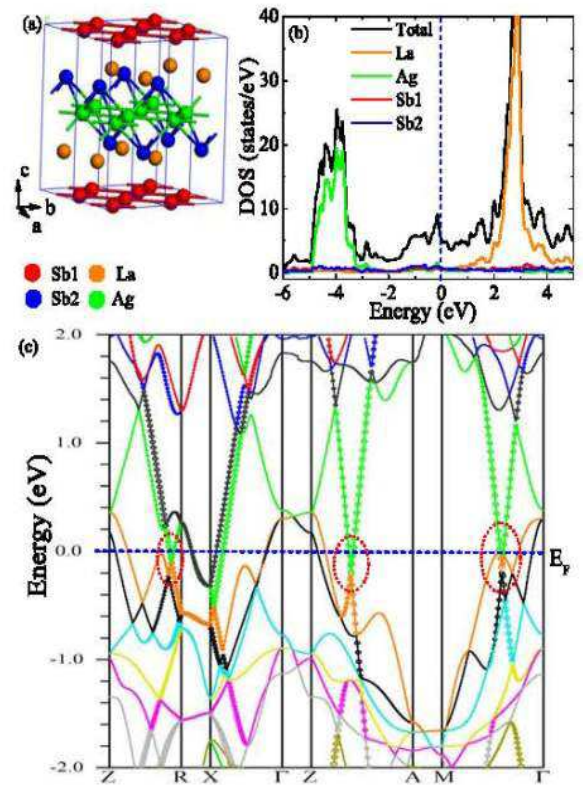}
\caption{(a) The crystal structure of LaAgSb$_2$. Sb atoms in 2D square nets (Sb1) are shows by red balls. Ag atoms are denoted by green balls. Sb atoms in AgSb$_4$ tetrahedrons (Sb2) are denoted by blue balls. La atoms are denoted by orange balls. Blue lines define the unit cell.(c) The total density of states (DOS) (black line) and local DOS from La, Ag, Sb1 and Sb2 in LaAgSb$_2$. The dotted line indicates the position of the Fermi energy. (d) The band structure for LaAgSb$_2$. The lines with open circles indicates the bands with $p_{x,y}$ character. The line at energy=0 indicates the position of the Fermi level and the red circles indicate the position of Dirac-cone-like points.}
\end{figure}

\begin{figure}[tbp]
\includegraphics[scale=1] {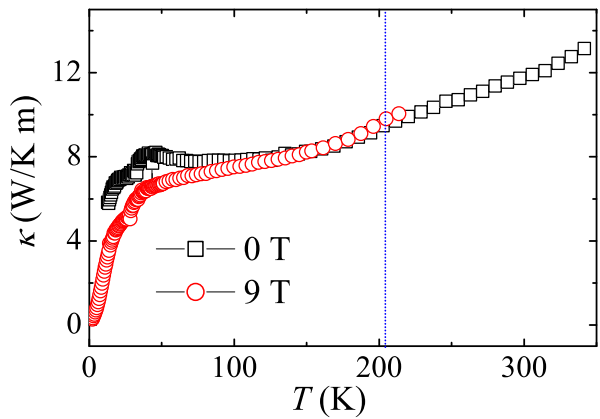}
\caption{In-plane thermal conductivity $\kappa(T)$ of LaAgSb$_2$ single crystal as a function of temperature in 0 T and 9 T magnetic field respectively. The blue line indicates the sharp anomaly at $\sim 210$ K.}
\end{figure}

\begin{figure}[tbp]
\includegraphics[scale=1] {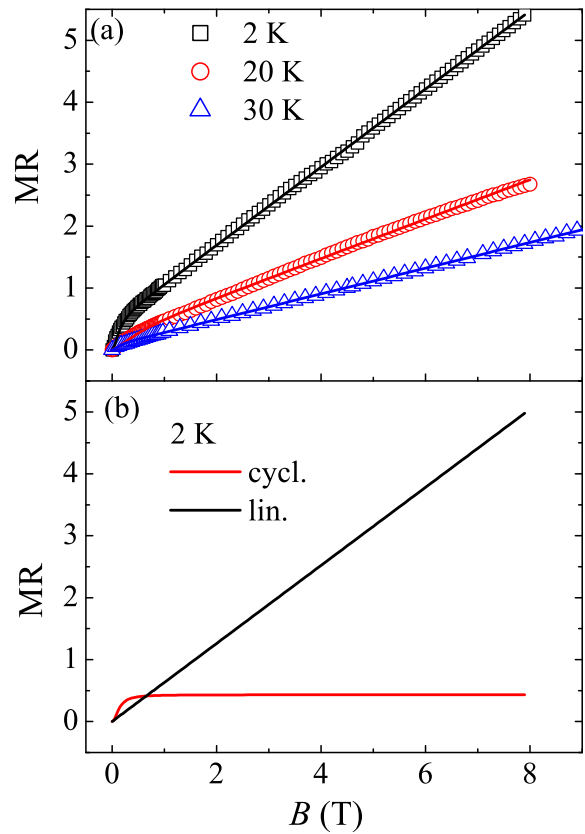}
\caption{(a) The magnetic field ($B$) dependence of the in-plane magnetoresistance MR at different temperatures from 2 K to 30 K. The open symbols are the experimental data and solid lines are the fitting results using equation (3). (b) The linear and cyclotronic contributions to the magnetoresistance at 2 K derived from fitting using equation (3).}
\end{figure}

LaAgSb$_2$ exhibits very large linear magnetoresistance. Below 30 K, the in-plane magnetoresistance MR=$(\rho_{ab}(B)-\rho_{ab}(0))/\rho_{ab}(0)$ exhibits a sharp dip at low field and then increases nearly linearly with increasing field at higher field ($>1$ T), as shown in Fig. 3(a). At higher temperature ($>40$ K), the dip at low field disappears (Fig. 4(a) and (b)). Below 1 T at 2 K, $d$MR$/dB$ is proportional to $B$ (as shown by lines in low-field regions in Fig. 4(c)), indicating the semiclassical quadratic field dependent MR ($\sim A_2B^2$). But above a characteristic field $B^*$, $d$MR$/dB$ deviates from the semiclassical behavior and saturates to a much reduced slope (as shown by lines in the high-field region in Fig. 4(c)). This indicates that the MR for $B>B^*$ is dominated by a linear field dependence plus a very small quadratic term (MR$=A_1B+O(B^2)$).  With increasing temperature, the magnetoresistance is gradually suppressed and the cross over field $B^*$ increases rapidly (Fig. 4(a)). Above 200 K, Linear MR becomes invisible in our magnetic field range (0$\sim$ 9 T) (Fig. 4(b)) and the MR in the whole field range is quadratic (Fig. 4(d)).

\begin{figure}[tbp]
\includegraphics[scale=0.8] {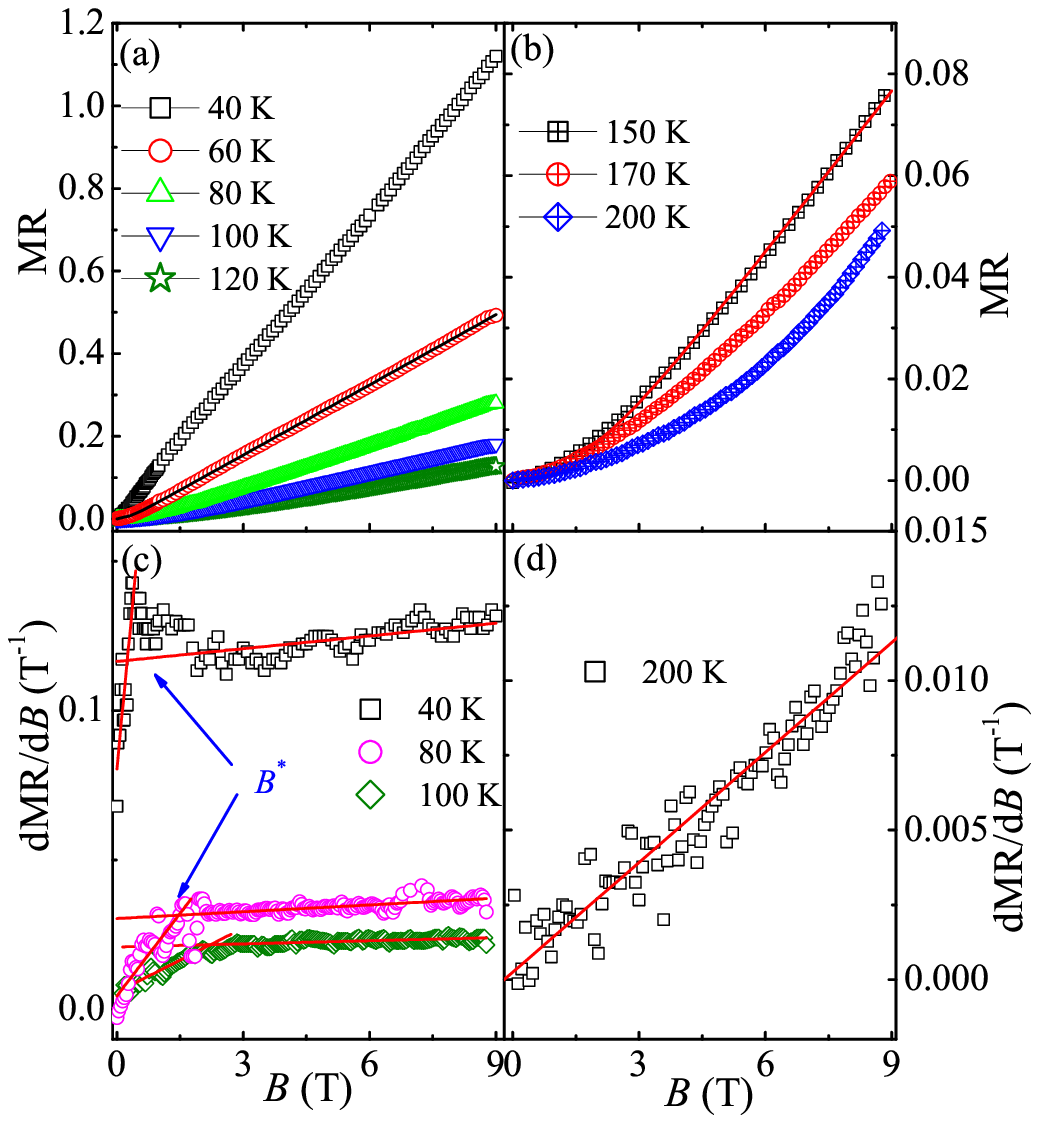}
\caption{(Color online) (a) and (b) The magnetic field ($B$) dependence of the in-plane magnetoresistance MR at different
temperatures above 40 K. (c) and (d) The field derivative of in-plane MR, $d$MR$/dB$, as a function of field (B) at different temperature
respectively. The red lines in high field regions were fitting results using MR $=A_1B+O(B^2)$ and the lines in low field regions using MR $=A_2B^2$.}
\end{figure}

Single band semiclassical transport gives that magnetoresistance scales as MR $=f(B\tau)=F(B/\rho_0)$ with the assumption of the single scattering time $\tau$, i.e. $1/\tau(T)\propto\rho_0(T)$, where $\rho_0(T)$ is the zero-field resistivity.\cite{mr3,kohler} The linear unsaturated magnetoresistance of LaAgSb$_2$ clearly deviates from semiclassical transport and also violates Kohler's scaling particularly in the high field region (Fig. 5(a)), indicating multiband or quantum effects.

In metals with two types of carriers (holes and electrons), semiclassical transport and the cyclotronic motion gives
\begin{eqnarray}
\frac{\rho(H)-\rho(0)}{\rho(0)} = \frac{\sigma_h\sigma_e(\mu_h+\mu_e)B^2}{(\sigma_h+\sigma_e)^2+(\sigma_h\mu_e+\sigma_e\mu_h)^2B^2},
\end{eqnarray}
where $\sigma_e (\sigma_h)$ and $\mu_e (\mu_h)$ are the electronic conductivity and mobility for electrons (holes) respectively. This formula usually results in the quadratic field-dependent MR in the low field range and saturated MR in the high field.\cite{mr3,lafeaso}

\begin{figure}[tbp]
\includegraphics [scale=0.95]{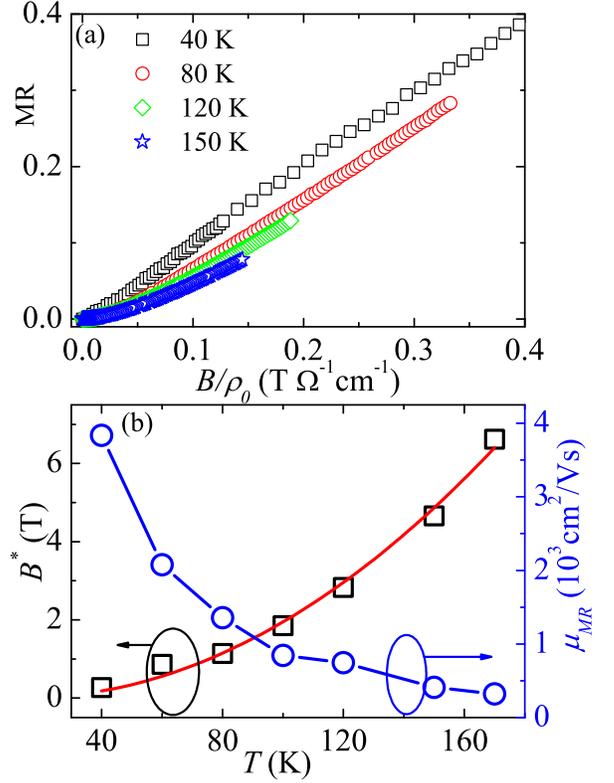}
\caption{(Color online) (a) The Kohler's plots for MR at different temperatures. (b)Temperature dependence of the critical field $B^*$ (black
squares) and the effective MR mobility $\mu_{MR}$ (blue circles) extracted from the weak-field MR. The red solid line is the fitting results of $B^*$ using $B^*=\frac{1}{2e\hbar
v_F^2}(E_F+k_BT)^2$.}
\end{figure}

Hitherto there are three possible mechanisms for linear MR. In the single crystal with open Fermi surface (such as quasi-one-dimension Fermi surface sheet), along the open orbits it will give $B^2$-dependent MR in very high field and the linear MR will possibly be observed in polycrystalline sample because of average effect.\cite{quantummr,mr3} This mechanism obviously does not attribute to our results since our sample is high quality single crystal. One of the other mechanisms for linear MR is the quantum limit in high magnetic field.\cite{quantumtransport,quantummr} Application of strong magnetic field ($B$) results in quantized Landau levels $E_n$ (LLs). When the field is very large and the difference between the zeroth and first Landau level $\triangle_{LL}$ exceeds the Fermi energy $E_F$ and the thermal fluctuation $k_BT$, all carriers only occupy the lowest LL and a large linear MR could be expected:
\begin{eqnarray}
MR = \frac{1}{2\pi}\left(\frac{e^2}{\varepsilon_{\infty}\hbar v_F}\right)^2\frac{N_i}{en^2}B\ln(\varepsilon_{\infty}),
\end{eqnarray}
where $N_i$ is the density of scattering centers, $n$ is the carrier density, $v_F$ is the Fermi velocity at the Dirac cones, and $\varepsilon_{\infty}$ is the high frequency dielectric constant.\cite{quantumtransport,quantummr} For electrons with conventional parabolic bands, $\triangle_{LL}=\frac{e\hbar B}{m^*}$ and its evolution with field is very slow because of the large effective mass $m^*$. Then it is impossible to observe quantum limit and linear MR behavior in the moderate field range (below 9 T). Unusual linear magnetoresistance in low temperature range was identified in some materials hosting Dirac fermions with linear energy dispersion, such as Ag$_{2-\delta}$(Te/Se), topological insulators and BaFe$_2$As$_2$.\cite{qt1,qt2,qt3,qt4,qt5} For Dirac states with linear energy dispersion, the energy splitting between the lowest and $1^{st}$ LLs is described by $\triangle_{LL}=\pm v_F\sqrt{2e\hbar B}$ where $v_F$ is the Fermi velocity. It increases rapidly with field because of the large Fermi velocity of Dirac fermions. Hence, in Dirac materials the quantum limit and quantum transport can be achieved in low field region.

Large linear MR was also observed recently in SrMnBi$_2$. In SrMnBi$_2$, highly anisotropic Dirac states were identified where linear energy dispersion originates from the crossing of two Bi $6p_{x,y}$ bands in the double-sized Bi square nets. The crystal structure of LaAgSb$_2$ has quasi-two-dimensional Sb layers similar to the double-sized Bi square nets in SrMnBi$_2$. It could be expected that the Sb layers in LaAgSb$_2$ can also host Dirac fermions. The band structure in Fig. 1(d) shows two nearly linear narrow bands crossing the Fermi level along $Z-A$ and $\Gamma-M$ directions. The Fermi level is located very close to the Dirac-cone-like points. Quantum oscillation experiments revealed the hollow cylindrical Fermi surface in LaAgSb$_2$ and very small effective mass ($m^*\sim 0.16m_0$ where $m_0$ is the mass of bare electron).\cite{oscillation1} These results suggest that the unusual linear MR would originate from the quantum limit of Dirac fermions. In Ref.[31], theoretical calculation predicted a very small pocket which has 20 T oscillation frequency and very small effective mass $\sim$ 0.06$m_e$, although it was not observed in the experiment. Our first principle result is consistent with this. Besides, another quantum oscillation measurement \cite{oscillation2} reports even smaller SdH oscillation frequencies $\sim$ 10 T and 80 T in LaAgSb$_2$. These indicate possible small Dirac pockets in the system. Most likely the quantum limit and linear MR is related to these very small pockets.

Taking the linear MR induced by quantum limit into account and combing Eqn.(1) and Eqn.(2), MR in LaAgSb$_2$ can be described very well by
\begin{eqnarray}
MR=\frac{\alpha B^2}{\beta+B^2}+\gamma|B|,
\end{eqnarray}
with $\alpha, \beta, \gamma$ as the fitting parameters.\cite{mr3,quantummr,lafeaso} The fitting curves for low temperature ($<$ 30 K) and higher temperature ($>$ 30 K) are shown in Fig. 3(a) and Fig. 4(a), respectively. Fig. 3(b) gives the MR contribution from cyclotron motion and quantum linear MR at 2 K, respectively. The cyclotron MR contribution has very low saturation value ($\sim 0.03$) and saturates at very small magnetic field ($\sim 0.4$ T). The critical condition to achieve quantum limit at finite temperature is $\Delta_{LL}=E_F+k_BT$ and then the critical field $B^*$ for Dirac fermions is $B^*=\frac{1}{2e\hbar v_F^2}(E_F+k_BT)^2$.\cite{qt3} For conventional electron gas with parabolic bands $B^*$ is proportional to temperature. The temperature dependence of critical field $B^*$ in LaAgSb$_2$ clearly deviates from the linear relationship and can be well fitted by $B^*=\frac{1}{2e\hbar v_F^2}(E_F+k_BT)^2$, as shown in Fig. 5(b). The fitting gives a large Fermi velocity $v_F\sim1.46\times 10^5$ ms$^{-1}$. The mobility of the system can be inferred from the semiclassical transport behavior in low field region. For a multiband system, the coefficient of the low-field semiclassical $B^2$ quadratic term, $A_2$, is related to the the effective electron and hole conductivity ($\sigma_e, \sigma_h$) and mobility ($\mu_e,\mu_h$) through $\sqrt{A_2}=\frac{\sqrt{\sigma_e\sigma_h}}{\sigma_e+\sigma_h}(\mu_e+\mu_h)=\mu_{MR}=\sqrt{A_2}$, which is smaller than the average mobility of carriers $\mu_{ave}=\frac{\mu_e+\mu_h}{2}$ and gives an estimate of the lower bound to the latter.\cite{qt3,qt4} Fig. 5(b) shows the dependence of $\mu_{MR}$ on the temperature. At 2 K, the value of $\mu_{MR}$ is about 4000 cm$^2$/Vs in LaAgSb$_2$ which is larger than the values in conventional metals and semiconductors. The parabolic temperature dependence of $B^*$ and the large $\mu_{MR}$ confirms the existence of Dirac fermion states in LaAgSb$_2$. Another possible reason of the evolution of the crossover field $B^*$ is the dependence of the carrier density and mobility.  Although there is temperature evolution of the carrier density and mobility, Dirac fermions dominate the MR behavior and the large linear MR should be only due to the quantum limit of Dirac fermions. The temperature evolution of carriers should only influence the curve shape in the semiclassical transport region and the magnitude of the quantum linear MR since the quantum linear MR only depends on the Dirac fermions density. In fact the decrease of MR with temperature increasing should be attributed to the decrease of Dirac carriers. But the crossover point from semiclassical transport region to quantum transport region should not be influenced by this temperature evolution of carrier density/mobility but by the thermal fluctuation smearing out the LL splitting.

Another possibility for linear MR is the breakdown of weak-field magnetotransport at a simple density-wave quantum critical point.\cite{qcp1,qcp2} Quasi-linear MR was also found in Sr$_2$RuO$_4$ \cite{srruo} and Ca$_3$Ru$_2$O$_7$.\cite{qcp1} Both of them are argued to be a small gap density wave system with quasi-2D Fermi surface. Theoretical analysis pointed out that at a simple density-wave quantum critical point the weak-field regime of magnetotransport collapses to zero-field with the size of gap.\cite{qcp1,qcp2} LaAgSb$_2$ also exhibits CDW transition at $\sim 200$ K and the linear MR disappears around the CDW temperature. So the linear MR or the Dirac fermions should be related to the CDW order. Similar phenomena was observed in BaFe$_2$As$_2$, in which the linear MR disappears above the spin density wave (SDW) temperature and the formation of Dirac fermions was attributed to the nodes of the SDW gap by complex zone folding in bands with different parities.\cite{qt4,qt5} The detail mechanism of linear MR, the existence of Dirac fermion, as well as the relationship between Dirac fermions and CDW in LaAgSb$_2$ deserve further study using direct methods such as the angular resolved photoelectron spectroscopy (ARPES).

\begin{figure}[tbp]
\includegraphics [scale=0.75]{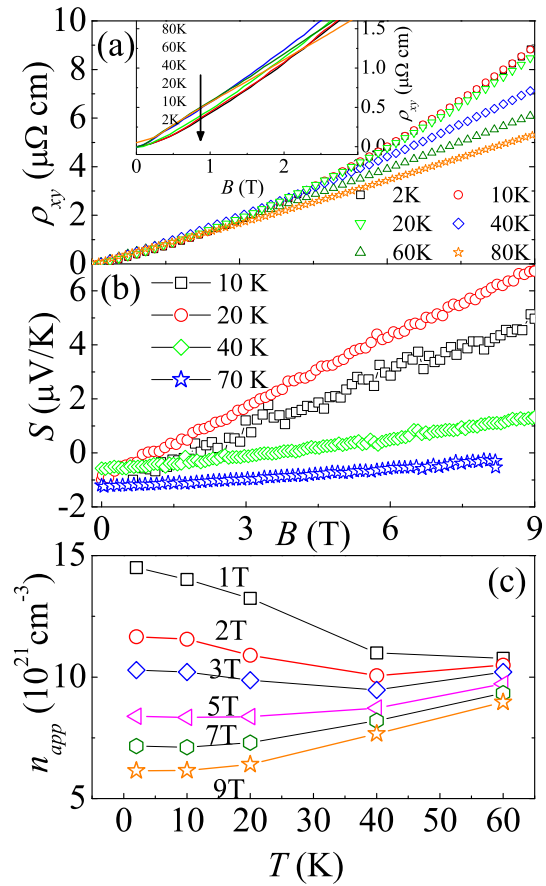}
\caption{(Color online) The magnetic field ($B$) dependence of the Hall resistivity $\rho_{xy}$ (a) and Seebeck coefficient $S$ (b) at different
temperatures. Note the sign change at $\sim 2$ T in $S(B)$ at temperature below 70 K. The  inset in (a) shows the Hall resistivity in field below 3 T. (c) The temperature dependence of the apparent carrier density as derived from Hall data in different fields.}
\end{figure}

\begin{figure}[tbp]
\includegraphics [scale=1]{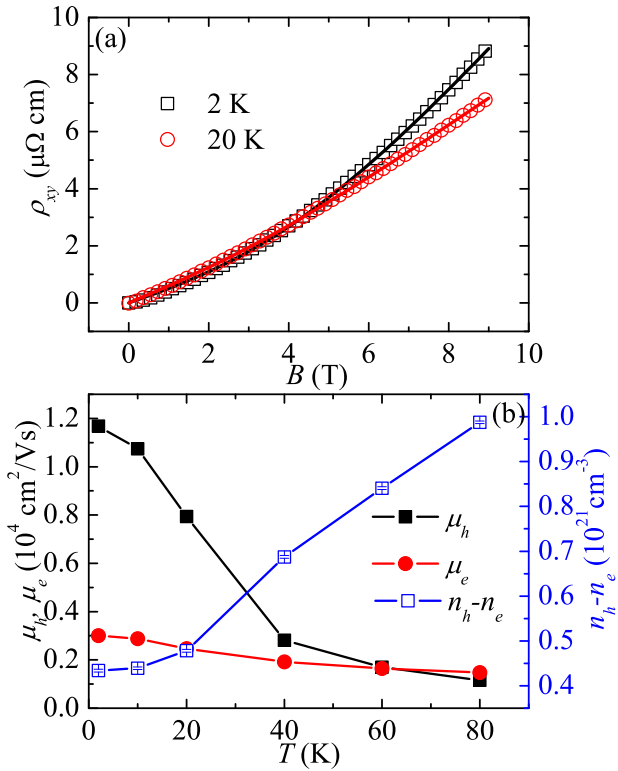}
\caption{(Color online) The fitting results for themagnetic field ($B$) dependence of the Hall resistivity $\rho_{xy}$ at 2 K and 20 K respectively. (b) The temperature dependent carrier mobility $\mu_h, \mu_e$ and the carrier density $(n_h-n_e)$ derived from the Hall resistivity fitting.}
\end{figure}

The magnetic field also has significant influence on the thermal transport of LaAgSb$_2$ (Fig. 2(b) and (c)). Magnetic field suppresses thermal conductivity significantly below 200 K due to the large MR. Fig. 6(a) and (b) show the magnetic field dependence of the Hall resistivity $\rho_{xy}$ and Seebeck coefficient $S$ at different temperatures respectively.  The behavior in $\rho_{xy}$ is different from the classical Hall behavior. The positive $\rho_{xy}$ are not linear in field but quadratic. The Hall resistivity curves all cross at $\sim 2$ T. Below 2 T $\rho_{xy}$ increases with increase in temperature, while it decreases with increase in temperature above 2 T. This indicates the change in the apparent carrier density $n_{app}=B/(e\rho_{xy})$. Fig. 6(c) shows the temperature dependence of $n_{app}$ at several magnetic fields. At 2 K, the apparent carrier density at 1 T is reduced by a half in 9 T field, implying the suppression of the DOS. This also induces the suppression of thermal conductivity in field (Fig. 2). More interestingly, $n_{app}$ increases with increase in temperature below 2 T while it decreases with increase in temperature for fields larger than 2 T. In zero field, Seebeck coefficient is negative in the whole temperature range. For magnetic field dependence of $S$ (Fig. 6(b)), the absolute value decreases linearly with increase in magnetic field below 2 T. It becomes zero at about 2 T when temperature is below 30 K. Further increase in field induces the changes of the sign of Seebeck coefficient from negative to positive and then the magnitude increases with increase in field.

In a single band metal with diffusion mechanism and electron-type carriers, Seebeck coefficient is given by the Mott relatioship,
\begin{eqnarray}
S=-\frac{\pi^2k_B^2T}{3e}\frac{\partial \ln\sigma(\mu)}{\partial \mu},
\end{eqnarray}
where $\rho(\varepsilon)$ is the DOS, $\varepsilon_F$ is the Fermi energy, $k_B$ is the Boltzman constant and $e$ is the absolute value of electronic charge.\cite{TE} The electron contribution to Seebeck coefficient $S_e$ is usually negative while the hole contribution $S_h$ is always positive.\cite{TE} For a two-band metal comprising electron and hole bands, $S$ is expressed as
\begin{eqnarray}
S=\frac{\sigma_h|S_h|-\sigma_e|S_e|}{\sigma_h+\sigma_e},
\end{eqnarray}
where $\sigma_{e(h)}$  and $S_{e(h)}$ are the contributions of electrons (holes) to the electric conductivity and Seebeck coefficient, respectively. According to the classical expression for the Hall coefficient including both electron and hole type carriers,\cite{mr3,mr}
\begin{eqnarray}
&\rho _{xy}&/\mu _{0}H = R_{H} \nonumber \\
&=&\frac{1}{e}\frac{(\mu _{h}^{2}n_{h}-\mu _{e}^{2}n_{e})+(\mu _{h}\mu
_{e})^{2}(\mu _{0}H)^{2}(n_{h}-n_{e})}{(\mu _{e}n_{h}+\mu
_{h}n_{e})^{2}+(\mu _{h}\mu _{e})^{2}(\mu _{0}H)^{2}(n_{h}-n_{e})^{2}},
\end{eqnarray}
where $e$ is the electron charge, $n_{e(h)}$ and $\mu_{e(h)}$ represent the carrier concentrations and mobilities of the electrons (holes).
Once there are two carrier types present, the field dependence of $\rho
_{xy}(H)$ will become nonlinear. Moreover, eq. (2) gives $R_{H}=\frac{1}{e}\frac{(\mu _{h}^{2}n_{h}-\mu _{e}^{2}n_{e})}{(\mu _{e}n_{h}+\mu _{h}n_{e})^{2}}$,
when $\mu _{0}H\rightarrow $ 0, and $R_{H}=e^{-1}\cdot 1/(n_{h}-n_{e})$ when
$\mu _{0}H\rightarrow $ $\infty $.\cite{mr}

First principle band structure in Fig. 1(c) shows that the Fermi level is located just below the Dirac-cone-like point of the gapless linear bands, but there is also a wide band crossing the Fermi level at $Z$ point. Most likely the Dirac holes and the conventional electrons contribute to the transport simultaneously. The parabolic curves of $\rho_{xy}$ plus the opposite signs between $\rho_{xy}$ and $S$(T) in low field reflect the multiband effect. Hall resistivity $\rho_{xy}$ can be fitted very well by the two-band Hall coefficient Eqn.(6), as shown in Fig. 7(a) by two typical curves at 2 K and 20 K. Combined it with MR fitting using Eqn.(1), we derive some parameters including carrier mobility ($\mu_h, \mu_e$) and carrier density $n=n_h-n_e$, as shown in Fig. 7(b). The holes have much higher mobility ($\sim 10^4$ cm$^2$/Vs) than the electrons at 2 K, and with increasing temperature, the hole mobility $\mu_h$ decreases significantly, but the change in the electron mobility $\mu_e$ is negligible. The value of mobility at 40 K derived from Hall resistivity is consistent with the magnitude from MR fitting. Moreover, the carrier density is positive indicating the hole density is higher than electron density.

Electronic transport in the Hall channel is determined by the density and mobility of different carriers, according to Eqn.(6).\cite{mr3} The Dirac holes have much higher mobility and dominate the Hall resistivity, which make $\rho_{ab}$ of LaAgSb$_2$ positive. But Seebeck coefficient is proportional to the logarithmic derivative of the DOS at Fermi level and then inversely proportional to the DOS at Fermi level or carrier density.\cite{TE} Especially, for two-dimensional Dirac system with linear energy dispersion, $S\propto1/\sqrt{n}$ could be expected and was indeed observed in graphene.\cite{TEgraphene1,TEgraphene2} The electron density is smaller than the hole density according to the two-band analysis of Hall resistivity and electron band has smaller DOS in LaAgSb$_2$, so $|S_e|$ is larger than $|S_h|$ and in zero field Seebeck coefficient $S$ is negative. With increasing magnetic field the Dirac holes will occupy the zeroth LLs gradually and dominate the thermal transport behavior in the quantum limit since the Fermi level locates between the zeroth and first LLs and the DOS at the Fermi level is suppressed. The positive Hall resistivity and Seebeck coefficient confirm the dominant hole-like carriers in high field which induce the sign change in $S(B)$ at $\sim 2$ T.

\section{Conclusion}

In conclusion, we performed detailed magnetoresistance and magnetothermopower measurement in LaAgSb$_2$ single crystal. The in-plane transverse magnetoresistance exhibits a crossover at a critical field $B^*$ from semiclassical weak-field $B^2$ dependence to the high-field linear-field dependence. The temperature dependence of $B^*$ satisfies quadratic behavior. This combined with the first principle electronic structure indicates the possible existence of Dirac fermions with linear energy dispersion. The linear magnetoresistance originates from the quantum limit of the possible Dirac fermions or the breakdown of weak-field magnetotransport at CDW transition. The Hall resistivity is positive, but the Seebeck coefficient is negative in 0 T field. Analysis of Hall resistivity using two-band model reveals that Dirac holes have higher mobility and larger density then conventional electrons, and dominate the electronic transport. Magnetic field suppresses the apparent Hall carrier density, and induces the sign change of the Seebeck coefficient from negative to positive. These effects are attributed to the magnetic field suppression of the density of states at Fermi level originating from the quantum limit of the Dirac holes.

\begin{acknowledgments}
We than John Warren for help with SEM measurements. Work at Brookhaven is supported by the U.S. DOE under contract No. DE-AC02-98CH10886.
\end{acknowledgments}


\end{document}